\documentclass[aps,superscriptaddress,showpacs,reprint]{revtex4-1}\pdfoutput=1
\usepackage{amsmath,latexsym,amssymb,hyperref,graphicx}
\usepackage{color}
\usepackage{cancel}
\usepackage{hyperref}
\usepackage{url}

\begin{document}

\title{Isospin-Violating Dark Matter and Neutrinos From the Sun}
\author{Shao-Long Chen}
\affiliation{Institute of Particle Physics and Key Laboratory of Quark \& Lepton Physics, Central China Normal University, Wuhan 430079, China}
\author{Yue Zhang}
\affiliation{International Center for Theoretical Physics, Trieste 34014, Italy}
\date{\today}

\begin{abstract}
We study the indirect detection of dark matter through neutrino flux from their annihilation in the center of the Sun, 
in a class of theories where the dark matter-nucleon spin-independent interactions break the isospin symmetry.
We point out that, while the direct detection bounds with heavy targets like Xenon are weakened and reconciled with the positive signals in 
DAMA and CoGeNT experiments, the indirect detection using neutrino telescopes can impose a relatively stronger constraint
and brings tension to such explanation, if the annihilation is dominated by heavy quark 
or $\tau$-lepton final states. As a consequence, the qualified isospin violating dark matter candidate has to preferably
annihilate into light flavors.
\end{abstract}

\maketitle

\section{Introduction}
It is well established that Dark Matter (DM) dominates the matter in the universe, but the identity of DM remains unclear. 
The direct detection experiments aim to decode the DM non-gravitational interactions by observing the scattering of DM 
off detector materials~\cite{Goodman:1984dc}. Many efforts have been made to search for such events for decades. 
Two collaborations, DAMA~\cite{Bernabei:2008yi} and CoGeNT~\cite{Aalseth:2010vx}, have claimed the evidences for 
annual modulation in the differential event rate, which is a characteristic property due to the motion of the Earth 
around the Sun~\cite{Drukier:1986tm}.  The simplest explanation points to a low mass ${\mathcal O}(10)$GeV DM 
spin-independently (SI) elastic scattering off nucleon with cross sections around $(2-5)\times 10^{-4}$ pb. 
It, however, turns out that DAMA tends to favor a relatively larger cross section than CoGeNT does.
Moreover, they contradict with the null experiments CDMS~\cite{cdms} and XENON~\cite{xenon} results, which put most 
stringent constraints on the SI DM-nucleon scattering cross sections.    

In order to alleviate the tension between the CoGeNT, DAMA results and the constraints of CDMS and XENON, 
various theoretical attempts and solutions have been put forward~\cite{Kopp:2009qt, Chang:2010yk}. 
Among them, isospin-violating dark matter (IVDM)~\cite{Chang:2010yk, Kurylov:2003ra, Feng:2011vu, Frandsen:2011ts} 
draws a lot of interests.  
It was proposed that the DM particles might couple differently to the protons and neutrons. Under this generic assumption,
one therefore gains an additional degree of freedom, $f_{n}/f_{p}$, the ratio between the two couplings. If it satisfies 
$f_{n}/f_{p}=-Z/(A-Z)$ for a given nuclear isotope $(A, Z)$, the scattering amplitudes will interfere destructively and cancel each other. Therefore, the constraints from the corresponding isotope could be completely evaded .   

There are also huge experimental efforts to detect DM indirectly through the detection of secondary products of DM annihilation 
in the galaxy or astrophysical bodies~\cite{Jungman:1995df, Bertone:2004pz}. 
One promising way is to detect the high energetic neutrino signals resulting from the annihilation of DM 
that have been gravitationally captured by the Sun~\cite{Press:1985ug, Gould:1987ir}, using the neutrino telescopes on the Earth, 
such as Super-Kamiokande (Super-K)~\cite{Desai:2004pq}  and IceCube~\cite{Abbasi:2009uz}. The most severe constraints set on the low mass DM are given by Super-K data. In this paper, we study the neutrinos flux coming from annihilation of 
IVDM in the Sun.

Throughout the discussion, we will assume the DM particle is symmetric, namely it can have significant annihilation when 
the number density is higher. We will comment on the scenario of asymmetric DM case in the end of the paper.

\section{Consequences of isospin violation}

In this section, we discuss the general consequences of isospin-violating DM-nucleon interactions in various 
DM direct/indirect detection approaches.
We will focus on the class of spin-independent interactions.

For ground-based direct detection experimental target containing a certain element $i$ with nucleon and proton numbers $(A_{i}, Z_{i})$, 
the ratio of the isospin-violating (IV) cross section to isospin-conservative (IC) cross section is
\begin{eqnarray}
\frac{\sigma_i^{\rm IV}}{\sigma_i^{\rm IC}}\sim \frac{[Z_i + (A_i - Z_i) f_n/f_p]^2}{A_i^2}\,.
\end{eqnarray}
The phenomenologically favored ratio is found to be $f_n/f_p\approx -0.7$. Due to destructive interference in the amplitude, 
the direct detection rate gets reduced significantly. 
The suppression factor turns out to be about $10^{-4}$ for Xenon and $10^{-3}$ for Germanium, while it is about $10^{-2}$ for Sodium. 
This feature acts as the key factor to reconcile the results of DAMA, CoGeNT and XENON experiments.

On the other hand, the capture of DM in the Sun is dominated by light elements for low mass DM favored by 
CoGeNT and DAMA results, namely Helium for isospin-conserving case and Hydrogen for isospin-violating case. 
The contributions of heavier elements are suppressed by their small chemical abundance. Therefore, the suppression factor
for capture is 
\begin{eqnarray}\label{rateratio}
\frac{C_\odot^{\rm IV}}{C_\odot^{\rm IC}} \sim \frac{\mu_{\rm H}^2}{A_{\rm He}^2 \mu_{\rm He}^2}\,,
\end{eqnarray}
where $\mu_i = m_\chi m_i/(m_\chi + m_i)$ and $m_i$ is the mass for nucleus $i$.
Taking into account of the presence of different isotopes, we list the reduction factors in direct detection and solar capture rates 
in Table.~\ref{4}.

\begin{table}[t]
\begin{tabular}{|c|c|c|c|c|}
\hline
Element &  Xe &  Ge & Na & Solar capture \\
\hline
Suppression &  $1.3\times10^{-4}$  & $2.6\times10^{-3}$ & $1.3\times10^{-2}$ &  $4.0\times10^{-2}$ \\
\hline
\end{tabular}
\caption{The suppression factors in the direct detection experiments and solar capture process, with $f_n/f_p=-0.7$ and $m_\chi=10\,$GeV. }
\label{4}
\end{table}

The key observation from Table.~\ref{4} is the hierarchy in the suppression factors,
amongst which solar capture rate receives the weakest suppression from isospin violation.
For DM mass around 10\,GeV, the capture rate is reduced only by a factor of 0.04. 

Therefore, the indirect detection using the neutrino flux can give relatively stronger bounds on the DM-nucleon SI interactions, 
if the interactions are isospin violating.
This serves as the main point of this paper. In the next section, we illustrate this statement quantitatively.

\section{Indirect detection from the Sun}
Weakly interacting DM can be captured in astrophysical bodies like the Sun. The capture process usually happens due to the scattering between DM and the nuclei. As DM particles are accumulated near the core region of the Sun, there can be significant annihilation process whose rate is proportional to its squared number density,
\begin{eqnarray}\label{capture}
\frac{d N}{d t} = C_\odot - C_A N^2\ ,
\end{eqnarray}
where $C_\odot$ is the capture rate and $C_A$ is the annihilation rate of DM particles in the Sun. 
For simplicity we neglect the evaporation term, which could be important for DM lighter than 3-4\,GeV~\cite{Hooper:2008cf}.
Assuming that $C_\odot$ and $C_A$ do not depend on time, one can readily solve the DM number $N(t)$
\begin{eqnarray}
N(t) = \sqrt{\frac{C_\odot}{C_A}} \tanh \left( {\sqrt{C_\odot C_A}}\cdot t \right) \ .
\end{eqnarray}
If the time needed to reach equilibrium is much smaller than the age of the solar system, i.e., $1/\sqrt{C_\odot C_A}\ll t_{\odot}$,
the capture and annihilation processes are now in equilibrium and the two terms on the right-handed side of Eq.~(\ref{capture}) 
are balanced with each other. 

\subsection{Capture rate}

The capture rate of DM by element $i$ in the Sun can be calculated by~\cite{Wikstrom:2009kw, Kappl:2011kz}
\begin{eqnarray}\label{rate}
C_{\odot, i} &=& \sum_i 4\pi \int_{0}^{R_\odot} r^2 dr \frac{\rho_\chi \rho_{\odot, i}(r)}{2 m_\chi \mu_i^2} \sigma_i \int_0^{\infty} du \frac{f(u)}{u} \nonumber \\
&\times&\theta(E_{R, \rm max}-E_{R, \rm cap}) \int_{E_{R, \rm cap}}^{E_{R, \rm max}} d E_R F^{2}(E_R)\,,
\end{eqnarray}
where the sum $i$ goes over all the elements and the DM local density is chosen to be $\rho_\chi = 0.3\,$GeV/cm$^3$. 
The capture rate $C_{\odot, i}$ is proportional to the corresponding DM-nucleus scattering cross section.
We use the standard chemical composition of the Sun given in~\cite{Grevesse:1998bj} to calculate the capture rate of each element, 
with the atomic number up to 56.  $\rho(r) = \rho_0 e^{-B\cdot{r}/{R_\odot}}$ is the mass density function of the Sun,
where $B=10.098$, $\rho_0 = M_\odot/\left(\int_0^{R_\odot} 4\pi s^2 ds e^{-B\cdot{s}/{R_\odot}} \right)$ and $M(r) = \int_0^{r} 4\pi s^2 ds e^{-B\cdot{s}/{R_\odot}}$. The mass density profile for each element is 
\begin{eqnarray}
\rho_{\odot, i}(r) = \rho (r) \cdot n_i\,,
\end{eqnarray}
where $n_i$ is the mass fraction for a given element $i$ in the Sun.
The DM velocity distribution is taken as a standard Maxwell-Boltzmann form,
\begin{eqnarray}
\frac{f(u)}{u} = \frac{1}{\sqrt{\pi} v_\odot^2} \left( e^{-(u-v_\odot)^2/v_\odot^2} - e^{-(u+v_\odot)^2/v_\odot^2} \right)\,,
\end{eqnarray}
where $v_\odot=220\,{\rm km}\,{\rm s}^{-1}$. 

\begin{figure}[t]
  \centering
  \includegraphics[width=.52\columnwidth]{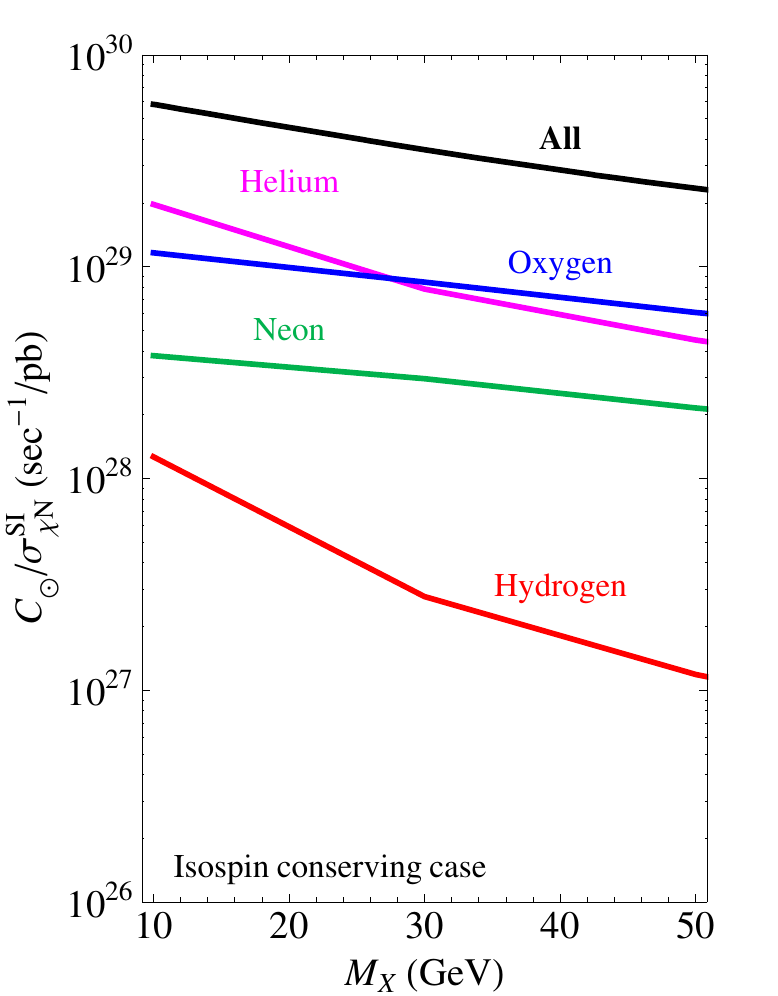}\hspace*{-0.2cm}
  \includegraphics[width=.52\columnwidth]{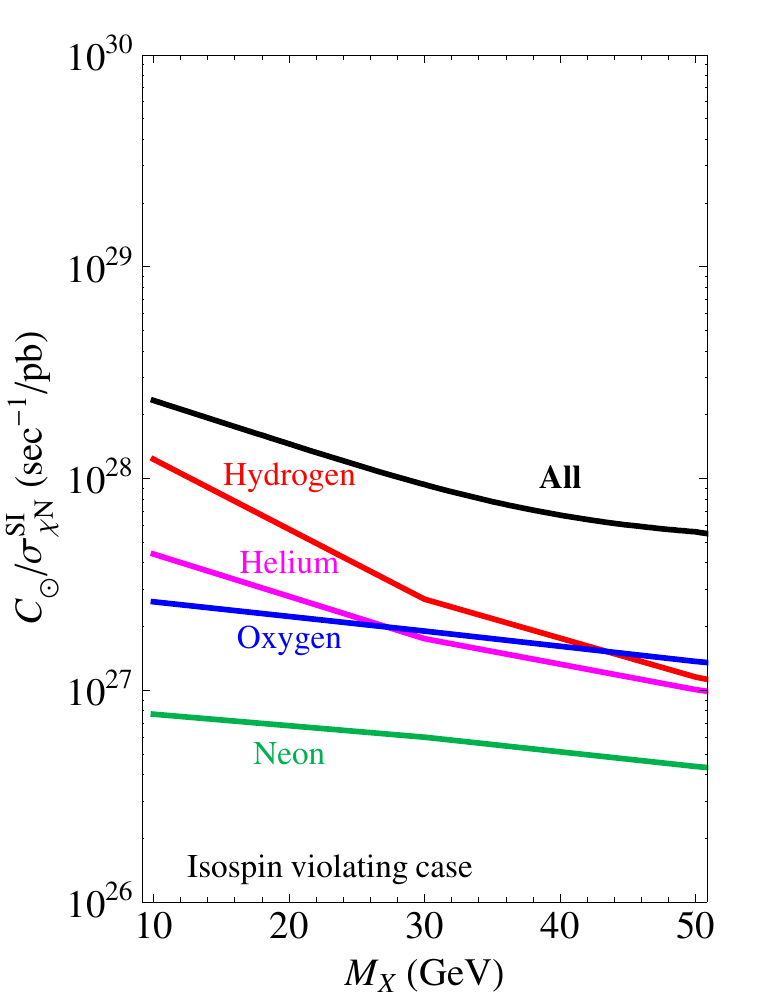}
  \caption{The solar capture rate per DM-nucleon spin-independent cross section for some representative elements in the isospin conserving and violating cases. For a light DM, Helium dominates 
  the contribution in the former case and Hydrogen in the latter.}\label{1}
\end{figure}

The nuclear form factor $F^2(E_{R})$ takes the Helm form 
\begin{eqnarray}
F^{2}(E_R) = e^{-E_R/E_i}\,,
\end{eqnarray}
where $E_i = 3/(2m_i R_i^2)$ and $R_i=(0.9 A^{1/3} + 0.3)\,$fm. 
For heavier nucleus, it is easier to deposit more energy in each scattering. Thus, the recoil energy integral is approximately proportional to the reduced mass squared $\mu_i^2$. This explains the additional factor of reduced mass ratio in Eq.~(\ref{rateratio}).

The lower limit for recoil energy integral in Eq.~(\ref{rate}) is the minimal energy transfer needed to capture the DM,
 $E_{R, \rm cap} = m_\chi u^2/2$; while the upper limit is the largest
energy transfer allowed by kinematics $E_{R, \rm max} = (2 \mu_i^2/m_i) \left(u^2 + v^2_{\rm esc}(r)\right)$.
The valid recoil energy integral must satisfy $E_{R, \rm cap}<E_{R, \rm max}$, so for the case $m_\chi\gg \mu_i$, the initial velocity $u$ at infinity must be small enough for the capture to happen.

The escape velocity of the Sun is given by
\begin{eqnarray}
v_{\rm esc} (r) \approx v_{\rm ctr}^2 - \frac{M(r)}{M_\odot} \left( v_{\rm ctr}^2 - v_{\rm surf}^2 \frac{R_\odot}{r} \right)\ ,
\end{eqnarray}
where for consistency $M_\odot \equiv M(R_\odot)  = 1.988\times10^{30}\,$kg, $R_\odot = 6.955\times10^8\,$m, $v_{\rm ctr} \equiv v_{\rm esc} (0)=1387.5\,$km/s and $v_{\rm surf} \equiv v_{\rm esc} (R_\odot)=617.5\,$km/s. 

\begin{figure}[t]
\centering
\includegraphics[width=.82\columnwidth]{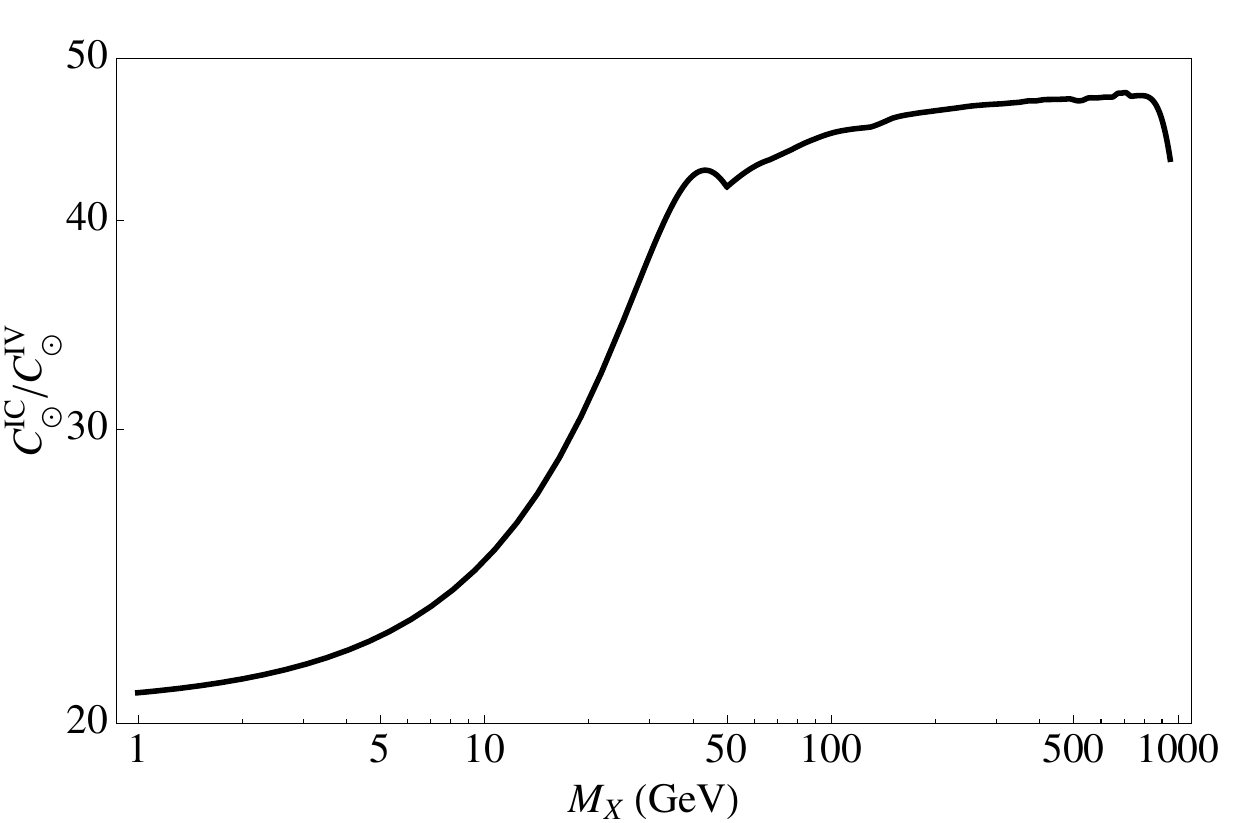}
\caption{The ratio of solar capture rates in isospin conserving and violating cases. For illustration,  
$f_n/f_p=-0.7$ is taken in isospin violating case.}\label{2}
\end{figure} 

\bigskip
The scattering cross section $\sigma_i$ in Eq.~(\ref{rate}) represents the strength of DM-nucleus $i$ interaction.
In the isospin-conserving case, the DM-nucleus scattering cross section is related to the cross section of the DM-proton by 
\begin{eqnarray}
\sigma_i = \sigma_{\chi p}^{\rm SI} A_i^2 \frac{\mu_i^2}{\mu_p^2}\ .
\end{eqnarray}
Compared with the Hydrogen (proton), the cross sections off heavy nucleus are enhanced by both $A_i^2$ and ${\mu_i^2}/{\mu_p^2}$~\cite{Agrawal:2010ax}. Of course, the capture rates for heavier elements are further suppressed by their abundance in the Sun.
Among all the relevant elements, as can be seen in Fig.~\ref{1}, the Helium plays the most dominant role in capturing the DM when the DM is lighter than 30\,GeV, while the Oxygen is most important for heavier DM.

On the other hand, in the isospin-violating case, the DM-nucleus scattering cross section is related to that between the DM and proton by
\begin{eqnarray}
\sigma_i = \sigma_{\chi p}^{\rm SI} [Z_i + (A_i-Z_i)f_n/f_p]^2 \frac{\mu_i^2}{\mu_p^2}\,.
\end{eqnarray}
The cross sections for heavy nuclei are suppressed by the destructive interference between protons and neutrons inside. 
Therefore it turns out that the Hydrogen is the dominant species to capture the DM with mass lower than 40 GeV, as shown in Fig.~\ref{1}.

In Fig.~\ref{2}, we plot the ratio of the total capture rates between isospin conserving and violating scenarios. 
As a rough estimate, the ratio is proportional to $A_{\rm He}^4 \rho_{\rm He}/\rho_{\rm H} \approx 20$ for $m_\chi\approx 10\,$GeV. 
As one can see in the plot, there is an upper bound $C_\odot^{\rm IC}/C_\odot^{\rm IV}\lesssim 50$ for $f_{n}/f_{p}=-0.7$.

\subsection{Annihilation and final state neutrinos}
The annihilation rate can be well approximated as
\begin{eqnarray}
C_A = \frac{\langle \sigma v \rangle}{V_{\rm eff}}\,,
\end{eqnarray}
where $V_{\rm eff}$ is the effective volume of the core of the Sun and found to be 
$V_{\rm eff}\approx 2.0 \times 10^{26}\,{\rm cm}^3 \left(\frac{1\,{\rm TeV}}{m_\chi}\right)^{3/2}$~\cite{Jungman:1995df}. 
There is in principle a competition between the capture and annihilation processes happening around the center of the Sun.
It has been shown that for the DAMA and CoGeNT favored region, the capture-annihilation equilibrium has been reached~\cite{Peter:2009mk}.
In fact, for fixed spin-independent interaction $\sigma_{\chi N}^{\rm SI}$ 
and annihilate rate $\langle \sigma v \rangle$, 
in isospin-violating scenario the processes reach equilibrium more quickly due to a smaller capture rate. After the capture and the annihilation 
processes become balanced, the flux of the annihilation process will be completely controlled by the capture rate. 

We mainly are interested in the final state neutrinos from the annihilation which can be detected by the neutrino telescopes 
such as the Super-K experiment. We use the results of Ref.~\cite{Cirelli:2005gh} to obtain the neutrino spectrum 
$\left({d N_{\nu_i}}/{d E_{\nu_i}}\right)_{F}$ per process, taking into account of hadronization, hadron stopping, neutrino 
absorption and assuming the effect of neutrino oscillation to the earth averages the three neutrino flavors~\cite{Jungman:1995df}. 
Here $F$ denotes the annihilation product of the DM. For light DM, the important final states are 
$\tau\bar\tau$, $b\bar b$ and $c\bar c$, which can further decay to neutrinos.
The neutrino flux when they arrive at the earth is then
$\left({d\Phi_{\nu_i}}/{d E_{\nu_i}}\right)_{F} = \left({d N_{\nu_i}}/{d E_{\nu_i}}\right)_{F} C_\odot ({1}/{4\pi R^2})$, where $R$ is the Sun-Earth distance. 

\subsection{Muon rate at Super-K}

\begin{figure}[t]
\begin{tabular}{@{}p{2.5in}p{4in}}
\includegraphics[width=0.73\columnwidth]{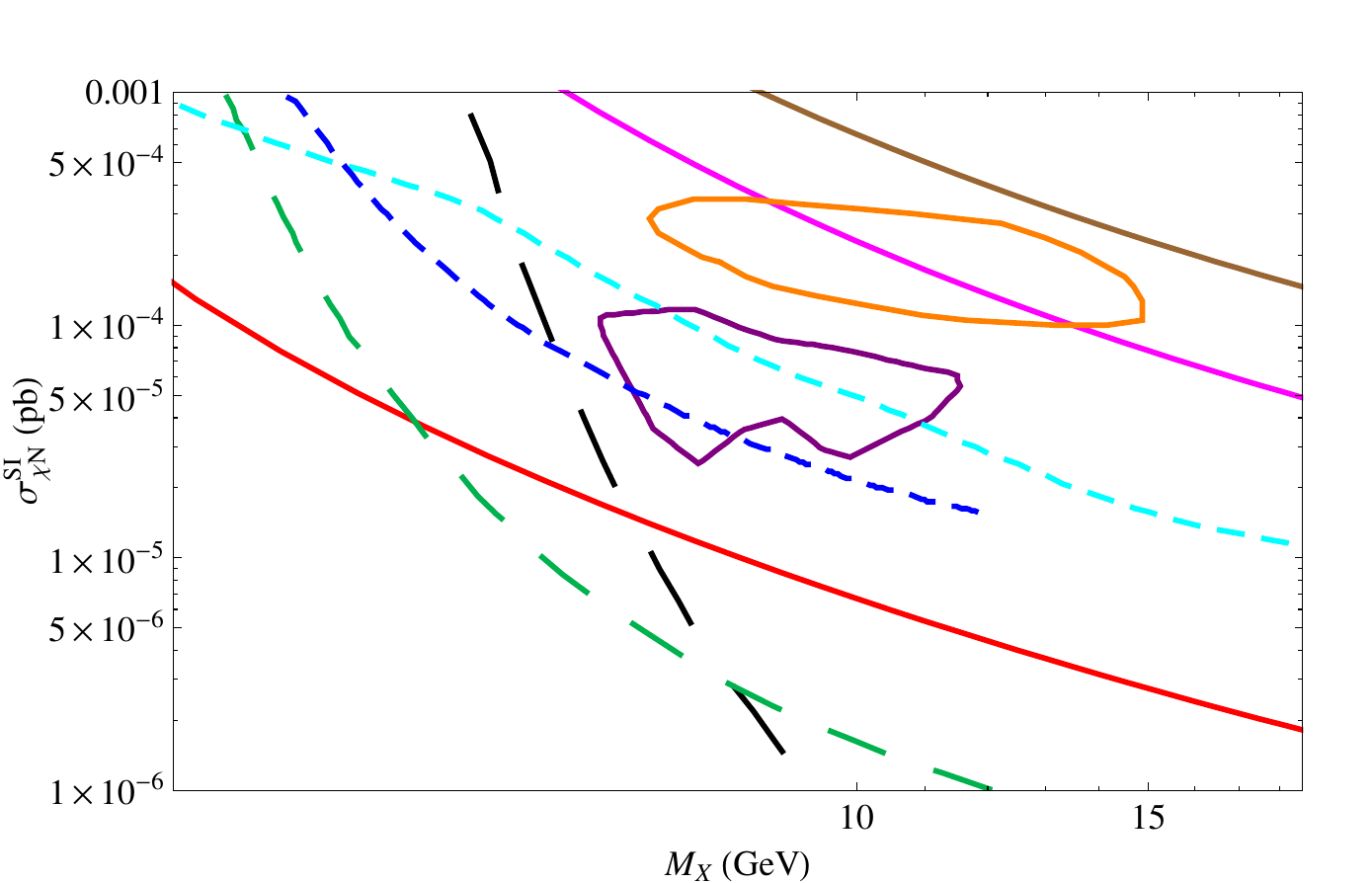} &\raisebox{-2.6cm}{\includegraphics[width=0.24\columnwidth]{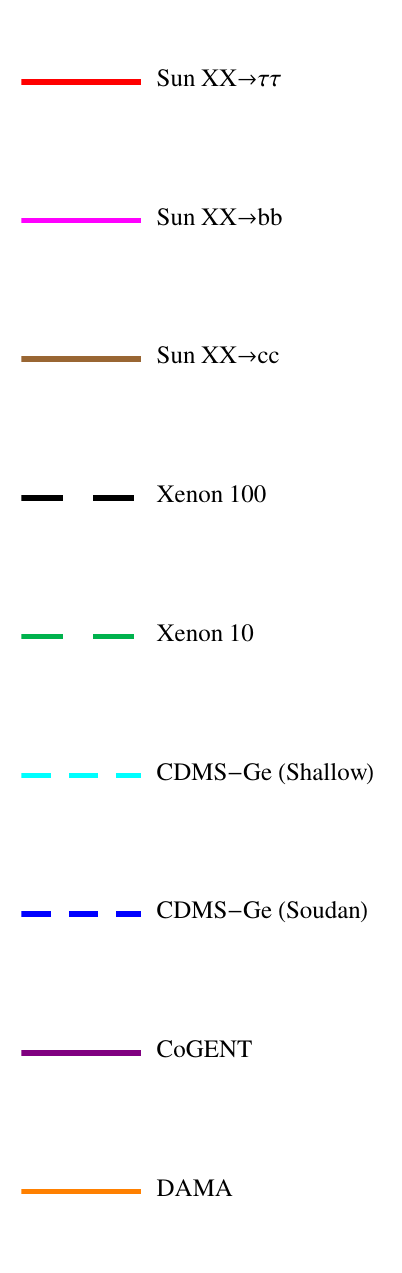}} \\
\vspace{-3cm}\includegraphics[width=0.73\columnwidth]{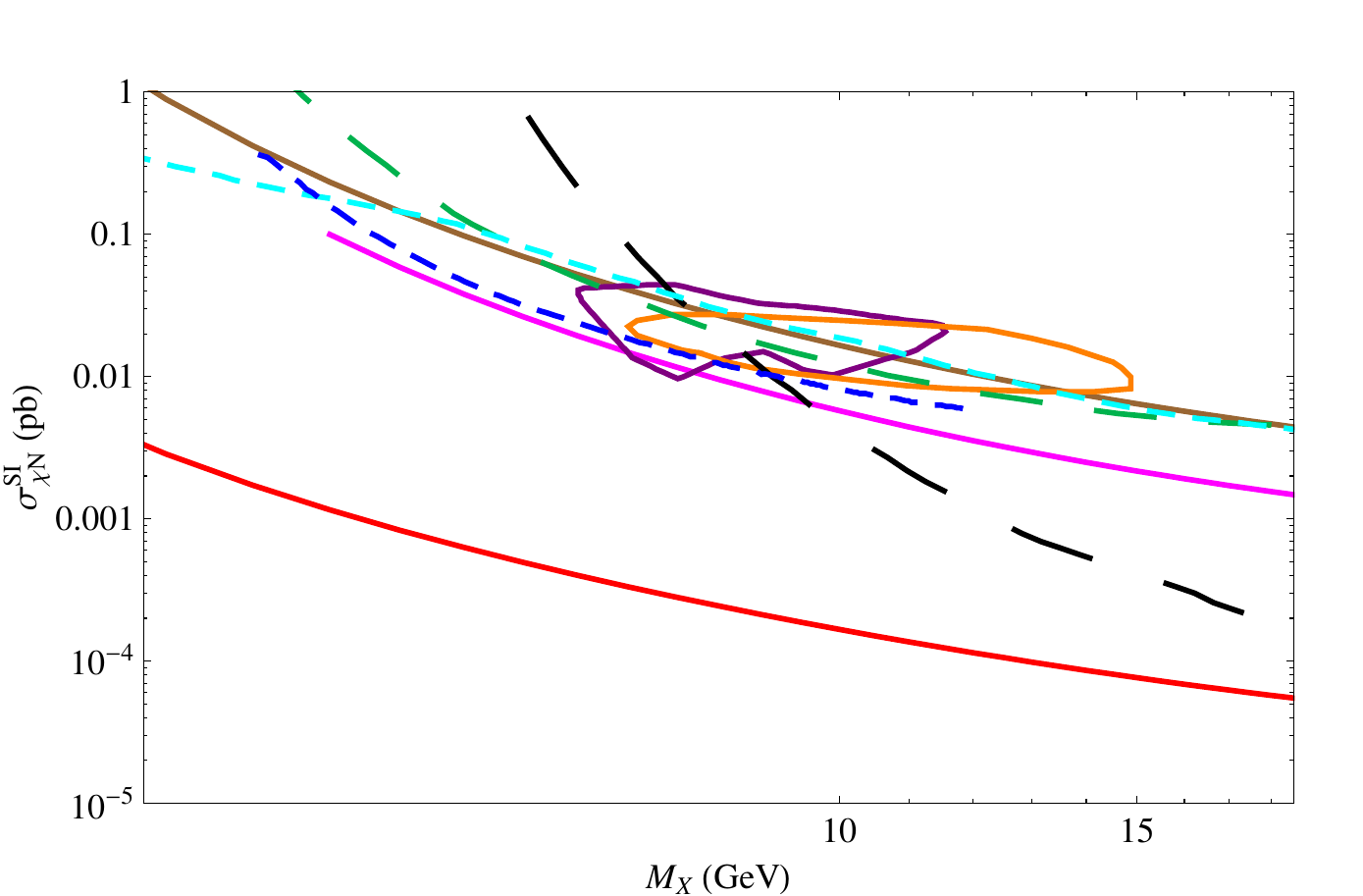} &
\end{tabular}
\caption{Positive signals from DAMA (orange circle) and CoGeNT (purple circle) in view of other direct detection experiments (dashed line) and 
indirect detection of neutrino flux (solid curves) from DM solar capture and annihilation, in isospin conserving (upper panel) and 
violating (lower panel) cases. In each panel, from up to down the solid curves represent annihilation to 
final states $c\bar c$, $b\bar b$ and $\tau\bar\tau$, assuming 100\% branching ratio.}\label{3}
\end{figure} 

The state-of-art technique to study the neutrino flux from cosmic rays is to observe up-going muons into the detector. 
These muons signals are generated by muon neutrino interacting with rocks as well as materials (e.g., water) inside the detector. 
Here we follow Refs.~\cite{Hooper:2008cf, Erkoca:2009by, Covi:2009xn} to calculate the muon rate generated in the presence of 
neutrino flux obtained in the previous subsection. 
\begin{eqnarray}
\Phi_{\mu} &=& \int_{E_\mu^{\rm th}}^{m_\chi} d E_\mu \int_{E_\mu}^{m_\chi} d E_{\nu_\mu} \frac{d \Phi_{\nu_\mu}}{d E_{\nu_\mu}}  \\ 
&&\times \left[ \frac{\rho}{m_p} \frac{d \sigma_\nu}{d E_\mu} (E_\mu, E_{\nu_\mu}) R_\mu (E_\mu, E_\mu^{\rm th}) \right] + ({\rm antineutrino})\,, \nonumber
\end{eqnarray}
where the square bracket represents the probability of muon being generated from charge-current weak interactions with cross section $\sigma_\nu$, and traveling though the average length $R_\mu$. $\rho$ is the mass density of the rocks or the water. The charged current interaction with nucleons can be written as
\begin{eqnarray}
\frac{d \sigma^{(p,n)}_\nu(E_\mu, E_{\nu_\mu})}{d E_\mu}  = \frac{2}{\pi} G_F^2 m_p \left( a_\nu^{(p,n)} + b_\nu^{(p,n)} \frac{E_\mu^2}{E_{\nu_\mu}^2}\right)\,, \nonumber \\
\end{eqnarray}
where $a_\nu^{(p)}=0.15$, $a_\nu^{(n)}=0.25$ and $b_\nu^{(p)}=0.04$, $b_\nu^{(n)}=0.06$. The interactions of anti-neutrinos are similar, 
but with parameters $a_{\bar \nu}^{(p)}=b_\nu^{(n)}$, $a_{\bar \nu}^{(n)}=b_\nu^{(p)}$ 
and $b_{\bar \nu}^{(p)}=a_\nu^{(n)}$, $b_{\bar \nu}^{(n)}=a_\nu^{(p)}$. 
The average length that muon travels before losing its energy below the detector threshold energy $E_\mu^{\rm th}$ is parametrized by
\begin{eqnarray}
R_\mu (E_\mu, E_\mu^{\rm th}) = \frac{1}{\beta \rho} \log \left( \frac{\alpha + \beta E_\mu}{\alpha + \beta E_\mu^{\rm th}} \right)\,,
\end{eqnarray}
where we have taken the parameters $\alpha=2.3\times 10^{-3}\,{\rm cm}^2{\rm g}^{-1} {\rm GeV}^{-1}$ 
and $\beta = 4.4\times 10^{-6}\,{\rm cm}^2{\rm g}^{-1}$ in the calculations.

The Super-K experiment~\cite{Desai:2004pq} measures the Cherenkov radiation of energetic muons generated in the charge-current interactions. The effective area of detection is around $A_{\rm eff} = 900\,{\rm m}^2$, and the $\tau = 1679.6$\,live days measurement allows at most 11 events other than originating from the atmosphere neutrino background~\cite{Hooper:2008cf} at 95\% confidence level. We use this as the upper bound on the number events $N_\mu = \Phi_{\mu} A_{\rm eff} \tau/2$ from DM annihilations in the Sun, where the factor $1/2$ accounts for the nighttime.

We have plotted the constraints on DM-nucleon cross section in Fig.~\ref{3}, including both direct detections and indirect detection via neutrinos from the Sun. We focus on the low mass DM region in light of the recent direct detection excitement.
As was noticed in~\cite{Feng:2011vu}, the positive signals from DAMA and CoGeNT can be reconciled by including isospin-violating DM-nucleon interactions. Isospin violation effect can also relieve the tension with the null results of XENON experiments, but cannot remove the constraints from CDMS which uses the same material as CoGeNT~\cite{Feng:2011vu}. 

An interesting finding is that the indirect detection with neutrinos from DM annihilation in the Sun imposes a stronger constraint if the annihilation final states are neutrino-rich, i.e., $\tau\bar \tau$ or $b\bar b$ (or marginally $c\bar c$), as shown by the solid curves in the Fig.~\ref{3}. The annihilation to light quarks or muon is still allowed, since they would lose most energy before decay, due to relatively longer lifetimes. 
Therefore, the qualified IVDM candidates should annihilate preferably into light flavors. 
This is a model-independent result, originating from the hierarchical nature characterized in reduction factors for different rates, as shown in Table.~\ref{4}.
It brings more challenge for the DM model building.
One possibility could be the portal bridges the DM sector only to the first generation fermions in the SM sector. 
The couplings to second generation quarks (like $c\bar c$) could then be induced but safely suppressed by the Cabibbo mixing angle.

A second message we can learn from Table~\ref{4} and Fig.~\ref{2} is, the suppression factor of the total capture rate 
is always larger than 1/50, still less suppressed compared to those in direct detection experiments. Therefore, even in the case of 
inelastic scattering where elements heavier than Helium are more likely to dominant~\cite{Nussinov:2009ft}, the indirect detection 
from the sun could still impose an important constraint.

Finally, we comment on the scenario where the DM particle is asymmetric. The asymmetric dark matter has been proposed in order to
understand why the baryonic matter and DM have similar relic densities~\cite{ADM}. In such scenario, due to the lack of its antiparticle, the 
dark matter cannot annihilate as they are accumulated inside the sun. If this happens, it is more difficult to have indirect detection signal of the 
asymmetric DM if it is completely stable and the above indirect detection bounds from the sun will no longer hold. However, if the dark matter 
self interaction is strong enough, the self capture could still play important roles and reveal its existence~\cite{Frandsen:2010yj}.

\section{Conclusion}
In summary, we studied the capture of low mass isospin-violating DM in the Sun and the corresponding neutrinos 
flux from their subsequent annihilation.  Low mass isospin-violating dark matter has been proposed to reconcile 
the annual modulation signals observed by DAMA and CoGeNT with the constraints put by the null direct-detection experiments. 
The isospin-violating effects make the scattering cross section of DM off certain isotope relatively suppressed, which therefore helps 
to solve the contradiction between positive signals with the null experiments.  
However, we find that the indirect detection of neutrino signals through the neutrino telescope Super-K sets stronger constraints on the 
DM-nucleon interactions and brings further tension to such explanation, if the DM particles annihilate into 
neutrino-rich final states, e.g., tau leptons or bottom quarks.

\section*{Acknowledgement}
We thank Prateek Agrawal and Neal Weiner for useful discussion and comments.
The work of S.L.C. is partially supported by the NSFC under Grant No. 11145001 
and the MOE of China with the program NCET-10-0424.

\end{document}